\documentclass[11pt, a4paper]{article}

\usepackage{amsmath}
\usepackage{graphicx}

\begin{document}

\title{  EM Algorithm for Estimation of the Offspring Distribution in Multitype Branching Processes with Terminal Types   
} 
\author{    Nina Daskalova 
}           

\maketitle

\begin{abstract}
{Multitype branching processes (MTBP) model branching structures, where the nodes of the resulting tree are objects of different types. One field of application of such models in biology is in studies of cell proliferation. A sampling scheme that appears frequently is observing the cell count in several independent colonies at discrete time points (sometimes only one). Thus, the process is not observable in the sense of the whole tree, but only as the "generation" at given moment in time, which consist of the number of cells of every type. This requires an EM-type algorithm to obtain a maximum likelihood (ML) estimation of the parameters of the branching process. A computational approach for obtaining such estimation of the offspring distribution is presented in the class of Markov branching processes with terminal types.
}
\end{abstract}


\section{Introduction}

Branching processes (BP) are stochastic models used in population dynamics. The theory of such processes could be found in a number of books (\cite{Harris}, \cite{Athreya}, \cite{Asmussen}), and application of branching processes in biology is discussed in \cite{Jagers}, \cite{Yakovlev1}, \cite{Kimmel}, \cite{Haccou}. Statistical inference in BP depends on the kind of observation available, whether the whole family tree has been observed, or only the generation sizes at given moments. Some estimators considering different sampling schemes could be found in \cite{Guttorp} and \cite{Yanev}. The problems get more complicated for multitype branching procesess (MTPB) where the particles are of different types (see \cite{Mode}). Yakovlev and Yanev \cite{Yakovlev} develop some statistical methods to obtain ML estimators for the offspring characteristics, based on observation on the relative frequencies of types at time $t$. Other approaches use simulation and Monte Carlo methods (see \cite{Hyrien}, \cite{Gonzalez}). 

When the entire tree was not observed, but only the objects existing at given moment, an Expectation Maximization (EM) algorithm could be used, regarding the tree as the hidden data. Guttorp \cite{Guttorp} presents an EM algorithm for the single-type process knowing generation sizes and in \cite{GonzalezSpringer2010} an EM algorithm is used for parametric estimation in a model of Y-linked gene in bisexual BP. Such algorithms exist for strictures, called Stochastic Context-free Grammars (SCFG). A number of sources point out the relation between MTBPs and SCFGs (see \cite{Sankoff}, \cite{Geman}). This relation has been used in previous work \cite{Daskalova} to propose a computational scheme for estimating the offspring distribution of MTBP using the Inside-Outside algorithm for SCFG (\cite{Lari}). A new method, related to this, but constructed entirely for BP will be presented here.

The EM algorithm specifies a sequence that is guaranteed to converge to the ML estimator under certain regularity conditions. The idea is to replace one difficult (sometimes impossible) maximization of the likelihood with a sequence of simpler maximization problems whose limit is the desired estimator. To define an EM algorithm two different likelihoods are considered -- for the "incomplete-data problem" and for "complete-data problem". When the incomplete-data likelihood is difficult to work with, the complete-data could be used in order to solve the problem. The conditions for convergence of the EM sequence to the incomplete-data ML estimator are known and should be considered when such an algorithm is designed. More about the theory and applications of the EM algorithm could be found in \cite{McLachlan}.

The paper is organized as follows. In Section 2 the model of MTBP with terminal types is introduced. Section 3 shows the derivation of an EM algorithm for estimating the offspring probabilities in general, and then proposes a recurrence scheme that could be used to ease the computations. A comprehensive example is given in Section 4 and the results of a simulation study are shown in Section 5.

\section{The Model}

A MTBP could be represented as $\mathbf{Z}(t) = (Z_1(t), Z_2(t), \ldots Z_d(t))$, where $Z_k(t)$ denotes the number of objects of type $T_k$ at time $t$, $k = 1, 2, \ldots d$. An individual of type $k$ has offspring of different types according to a $d$-variate distribution $p_k(x_1, x_2, \ldots, x_d)$ and every object evolves independently. If $t = 0, 1, 2, \ldots$, this is the Bienaym\'{e}-Galton-Watson (BGW) process. For the process with continuous time $t \in [0, \infty)$, define the \textit{embedded generation process} as follows (see \cite{Athreya}).
Let $\mathbf{Y}_n =$ \textit{number of objects in the $n$-th generation of} $\mathbf{Z}(t)$. If we take the sample tree $\pi$ and transform it to a tree $\pi'$ having all its branches of unit length but otherwise identical to $\pi$, then $\mathbf{Y}_n(\pi) = \mathbf{Z}_n(\pi')$, where $\mathbf{Z}_n$ is a BGW process. We call $\mathbf{Y}_n$ the embedded generation process for $\mathbf{Z}(t)$. The trees associated either with BGW process, or the embedded generation BGW process will be used to estimate the offspring probabilities.

Now we consider MTBP where certain \textit{terminal types} of objects, once created, neither die nor reproduce (see \cite{Sankoff}). If $T = \{T_1, T_2, \ldots, T_m\}$ is the set of non-terminal types and $T^T= \{T^T_1, T^T_2, \ldots, T^T_{d-m}\}$ is the set of terminal types, then an object of type $T_i$ produces offspring of any type and an object of type $T^T_j$ does not reproduce any more. Here each $T^T_i \in T^T$ constitutes a \textit{final group} (see \cite{Harris}). This way we model a situation where "dead" objects do not disappear, but are registered and present as "dead" through the succeeding generations. The process described above is reducible, because once transformed into a terminal type, an object remains in it's final group. For irreducible processes some statistical theory has been developed (see \cite{Badalbaev} for example), but for reducible ones statistical inference is case-dependent.

We are interested in estimation of the offspring probabilities. If the whole tree $\pi$ is observed the natural ML estimator for the offspring probabilities is 
\begin{equation}\label{ndaskalova:mle}
\hat{p}(T_v \rightarrow \mathcal{A}) = \frac{c(T_v \rightarrow \mathcal{A})}{c(T_v)},
\end{equation}
where $c(T_v)$ is the number of times a node of type $T_v$ appears in the tree $\pi$ and $c(T_v \rightarrow \mathcal{A})$ is the number of times a node of type $T_v$ produces offspring $\mathcal{A}$ in $\pi$.
It is not always possible to observe the whole tree though, often we have the following sampling scheme $\{\mathbf{Z}(0), \mathbf{Z}(t)\}$, for some $t > 0$. Let $\mathbf{Z}(0)$ consists of 1 object of some type. Suppose we are able to observe a number of independent instances of the process, meaning that they start with identical objects and reproduce according to the same offspring distribution. Such observational scheme is common in biological laboratory experiments. If $t$ is discrete $\mathbf{Z}(t)$ is the number of objects in the $t$-th generation. For continuous time it is a "generation" in the embedded generation BGW process. A simple illustration is presented in fig. 1 a)--c), where "alive" objects are grey, "dead" ones are white and numbers represent the different types.

\begin{figure}
\centering
\includegraphics[width=145mm]{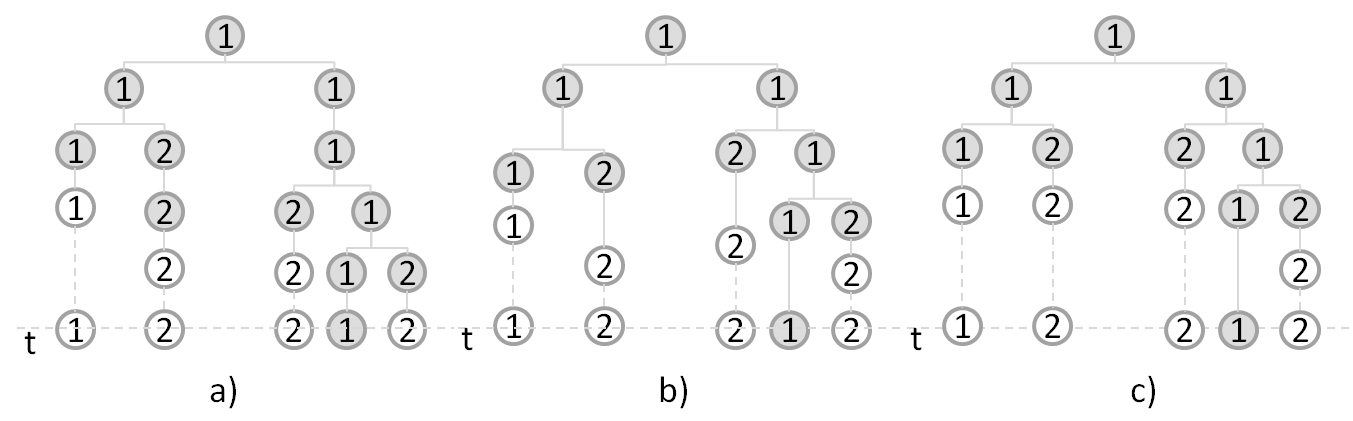}
\caption{A discrete time process a), continuous time process b) and its embedded process c).}
\end{figure}


Here the notion that "dead" objects present and could be observed in succeeding generations as terminal types is crucial. If "dead" particles disappeared somewhere inside the "hidden" tree structure, estimation would be impossible (see fig. 2 for an example).

\begin{figure}[h]
\centering
\includegraphics[width=85mm]{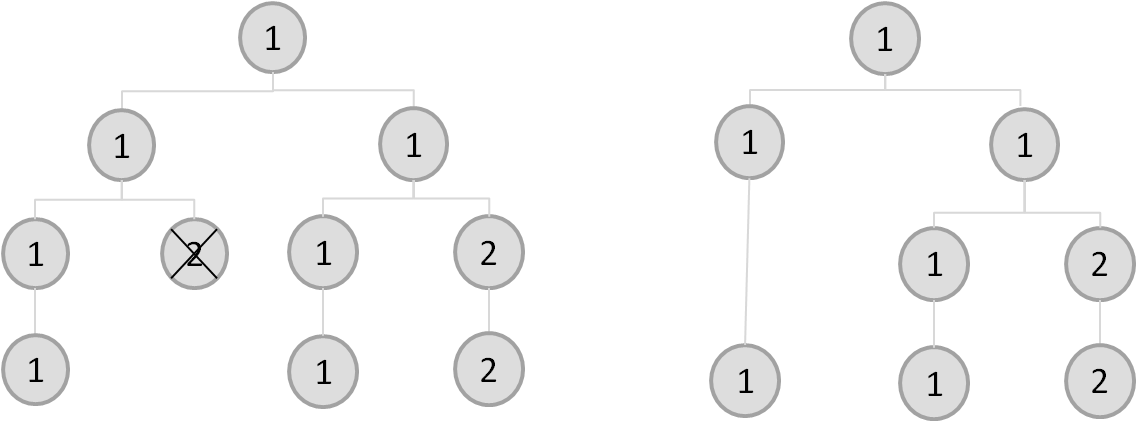}
\caption{Because of disappearing of a particle of type $2$ in the first tree information about the reproduction has been lost and we have the same observation as in the second tree.}
\label{disappear}
\end{figure}

\section{The EM Proposal}

\subsection{The EM Algorithm}

The EM algorithm was explained and given its name in a paper by Dempster, Laird, and Rubin  in 1977 \cite{Dempster}. It is a method for finding maximum likelihood estimates of parameters in statistical models, where the model depends on unobserved latent variables. Let a statistical model be determined by parameters $\theta$, $x$ is the observation and $Y$ is some "hidden" data, which determines the probability distribution of $x$. Then the density of the "complete" observation is $f(x,y|\theta)$ and the density  of the "incomplete" observation is the marginal one $g(x|\theta) = \int f(x,y|\theta)dy$. We can write the likelihoods and the conditional density of $Y$ given $x$ and $\theta$ this way: 
\begin{equation*}
L(\theta|x,y) = f(x,y|\theta), \quad L(\theta|x) = g(x|\theta), \quad and \quad k(y|\theta,x) = \frac{f(x,t|\theta)}{g(x|\theta)}.
\end{equation*}
The aim is to maximize the log likelihood 
\begin{equation*}\log L(\theta|x) = \log g(x|\theta) = \log L(\theta|x,y) - \log k(y|\theta,x).
\end{equation*}
As $y$ is not observed the right side is replaced with its expectation under $k(y|\theta',x)$:
\begin{equation*}
\log L(\theta|x) = E_{\theta'}[\log L(\theta|x,Y)] - E_{\theta'}[\log k(Y|\theta,x)].
\end{equation*}
Let $Q(\theta|\theta') = E_{\theta'}[\log L(\theta|x,Y)].$
It has been proved that
\begin{equation*}
\log L(\theta|x) - \log L(\theta^{(i)}|x) \geq Q(\theta|\theta^{(i)}) - Q(\theta^{(i)}|\theta^{(i)})
\end{equation*} 
with equality only if $\theta = \theta^{(i)}$, or if $k(Y|x,\theta^{(i)}) = k(Y|x,\theta)$ for some other $\theta \neq \theta^{(i)}$. Choosing $\theta^{(i+1)} = argmax_{\theta}Q(\theta|\theta^{(i)})$ will make the difference positive and so, the likelihood could only increase in each step. The \textit{Expectation Maximization Algorithm} is usually stated formally like this:

\textit{E-step:} Calculate function $Q(\theta|\theta^{(i)})$.

\textit{M-step:} Maximize $Q(\theta|\theta^{(i)})$ with respect to $\theta$.

The convergence of the EM sequence $\{\hat{\theta}^{(i)}\}$ depends on the form of the likelihood $L(\theta|x)$ and the expected likelihood $Q(\theta|\theta^{(i)})$ functions. The following condition, due to Wu (\cite{Wu}), guarantees convergence to a stationary point. It is presented here as cited in \cite{Casella}.

\textit{
If the expected complete-data log likelihood $E_{\theta'}[logL(\theta|x, Y)]$ is continuous both in $\theta$ and $\theta'$, then all limit points of an EM sequence $\{\hat{\theta}^{(i)}\}$ are stationary points of $L(\theta|x)$, and $L(\hat{\theta}^{(i)}|x)$ converges monotonically to $L(\hat{\theta}|x)$ for some stationary point $\hat{\theta}$.
}

\subsection{Derivation of an EM Algorithm for MTBP}

Let $x$ be the observed set of particles, $\pi$ is the unobserved tree structure and $\mathbf{p}$ is the set of parameters -- the offspring probabilities $p(T_v \rightarrow \mathcal{A})$ (the probability that a particle of type $T_v$ produces the set of particles $\mathcal{A}$). Then the likelihood of the "complete" observation is:
\begin{equation*}
L(\mathbf{p}|\pi,x) = P(\pi,x|\mathbf{p}) 
= \prod_{v,\mathcal{A}: T_v \rightarrow \mathcal{A}}p(T_v \rightarrow \mathcal{A})^{c(T_v \rightarrow \mathcal{A};\pi,x)},\end{equation*}
where $c$ is a counting function -- $c(T_v \rightarrow \mathcal{A};\pi,x)$ is the number of times a particle of type $T_v$ produces the set of particles $\mathcal{A}$ in the tree $\pi$, observing $x$.
The probability  of the "incomplete" observation is the marginal probability $P(x|\mathbf{p}) = \sum_{\pi}P(\pi,x|\mathbf{p})$.
For the EM algorithm we need to compute the function
\begin{flalign}\label{ndaskalova:Qfun}
\nonumber & Q(\mathbf{p}|\mathbf{p}^{(i)}) = E_{\mathbf{p}^{(i)}}(\log P(\pi,x|\mathbf{p})) = \sum_{\pi}P(\pi|x,\mathbf{p}^{(i)})\log P(\pi,x|\mathbf{p})\\  \nonumber
& = \sum_{\pi}P(\pi|x,\mathbf{p}^{(i)})\sum_{T_v \rightarrow \mathcal{A}}c(T_v \rightarrow \mathcal{A};\pi,x)\log p(T_v \rightarrow \mathcal{A})\\ \nonumber
& = \sum_{T_v \rightarrow \mathcal{A}}\sum_{\pi}P(\pi|x,\mathbf{p}^{(i)})c(T_v \rightarrow \mathcal{A};\pi,x)\log p(T_v \rightarrow \mathcal{A})\\ 
& = \sum_{T_v \rightarrow \mathcal{A}}E_{\mathbf{p}^{(i)}}c(T_v \rightarrow \mathcal{A})\log p(T_v \rightarrow \mathcal{A}).
\end{flalign}

We need to maximize the function (\ref{ndaskalova:Qfun}) under the condition $\sum_{\mathcal{A}}p(T_v \rightarrow \mathcal{A}) = 1$, $p(T_v \rightarrow \mathcal{A}) \geq 0$ for every $v$ and $A$.
The Lagrange method requires to introduce the function 
\begin{equation*}
\Phi(p) = \sum_{T_v \rightarrow \mathcal{A}}E_{\mathbf{p}^{(i)}}c(T_v \rightarrow \mathcal{A})\log p(T_v \rightarrow \mathcal{A}) +\lambda(1-\sum_{\mathcal{A}}p(T_v \rightarrow \mathcal{A})).
\end{equation*}
Taking partial derivatives with respect to $p(T_v \rightarrow \mathcal{A})$ and obtaining the Lagrangian multiplier
$\lambda = \sum_{\mathcal{A}}E_{\mathbf{p}^{(i)}}(T_v \rightarrow \mathcal{A})$, we get to the result that the re-estimating parameters are the normalized expected counts, which look like the ML estimators in the "complete" observation case (\ref{ndaskalova:mle}), but the observed counts are replaced with their expectations. 
\begin{equation}\label{ndaskalova:emest}
p^{(i+1)}(T_v \rightarrow \mathcal{A}) = \frac{E_{\mathbf{p}^{(i)}}c(T_v \rightarrow \mathcal{A})}{\sum_{\mathcal{A}}E_{\mathbf{p}^{(i)}}c(T_v \rightarrow \mathcal{A})} = \frac{E_{\mathbf{p}^{(i)}}c(T_v \rightarrow \mathcal{A})}{E_{\mathbf{p}^{(i)}}c(T_v)},
\end{equation}
where the expected number of times a particle of type $T_v$ appears in the tree $\pi$ is:
\begin{equation*}
E_{\mathbf{p}^{(i)}}c(T_v) = \sum_{\pi}P(\pi|x, \mathbf{p}^{(i)})c(T_v;\pi,x),
\end{equation*}
and the expected number of times a particle of type $T_v$ gives offspring $\mathcal{A}$ in the tree $\pi$ is:
\begin{equation*}
E_{\mathbf{p}^{(i)}}c(T_v \rightarrow \mathcal{A}) = \sum_{\pi}P(\pi|x, \mathbf{p}^{(i)})c(T_v \rightarrow \mathcal{A};\pi,x).
\end{equation*}

It is easy to check that in this case the convergence condition stated above is fulfilled. We consider the representation
\begin{equation*}
Q(\mathbf{p}|\mathbf{p}^{(i)}) = \sum_{T_v \rightarrow \mathcal{A}}\sum_{\pi}P(\pi|x,\mathbf{p}^{(i)})c(T_v \rightarrow \mathcal{A};\pi,x)\log p(T_v \rightarrow \mathcal{A}),
\end{equation*}
where $P(\pi|x,\mathbf{p}^{(i)})$ is a polynomial function of $p^{(i)}$-s -- the offspring probabilities, estimated on the $i$-th step. Then $Q(\mathbf{p}|\mathbf{p}^{(i)})$ is a sum of continuous functions in all the parameters $p$ and $p^{(i)}$, so it is also continuous. This way we are sure to converge to a stationary value, though it might not always be a global maximizer.  

It is a case of EM where the M-step is explicitly solved, so the computational effort will be on the E-step. The problem is that in general enumerating all possible trees $\pi$ is of exponential complexity. The method proposed below is aimed to reduce complexity.

\subsection{The Recurrence Scheme}

We have shown that the E-step of the algorithm consists of determining the expected number of times a given type $T_v$ or a given production $T_v \rightarrow \mathcal{A}$ appears in a tree $\pi$. A general method will be proposed here for computing these counts. The algorithm consists of three parts -- calculation of the inner probabilities, the outer probabilities and EM re-estimation, which are shown below.

Let us define the \textbf{inner} probability $\alpha({\mathbf I},v)$ of a subtree rooted at particle of type $T_v$ to produce outcome ${\mathbf I} = \{i_1, i_2, \ldots, i_d\}$ where $i_k$ is the number of objects of type $k$ (fig. 3). From the basic branching property of the process we get the following recurrence:
\begin{flalign*}
& \alpha({\mathbf I},v) = \sum_{\mathbf w} p(T_v \rightarrow \{T_{w_1}, \ldots, T_{w_k}\}) \sum_{{\mathbf I_1}+\ldots+{\mathbf I_k}={\mathbf I}} \alpha({\mathbf I_1},T_{w_1})\ldots\alpha({\mathbf I_k},T_{w_k})
\end{flalign*}
where ${\mathbf w} = \{T_{w_1}, \ldots, T_{w_k}\}$ are all possible sets of particles that $T_v$ can produce.

\begin{figure}[h]
\centering
\includegraphics[width=85mm]{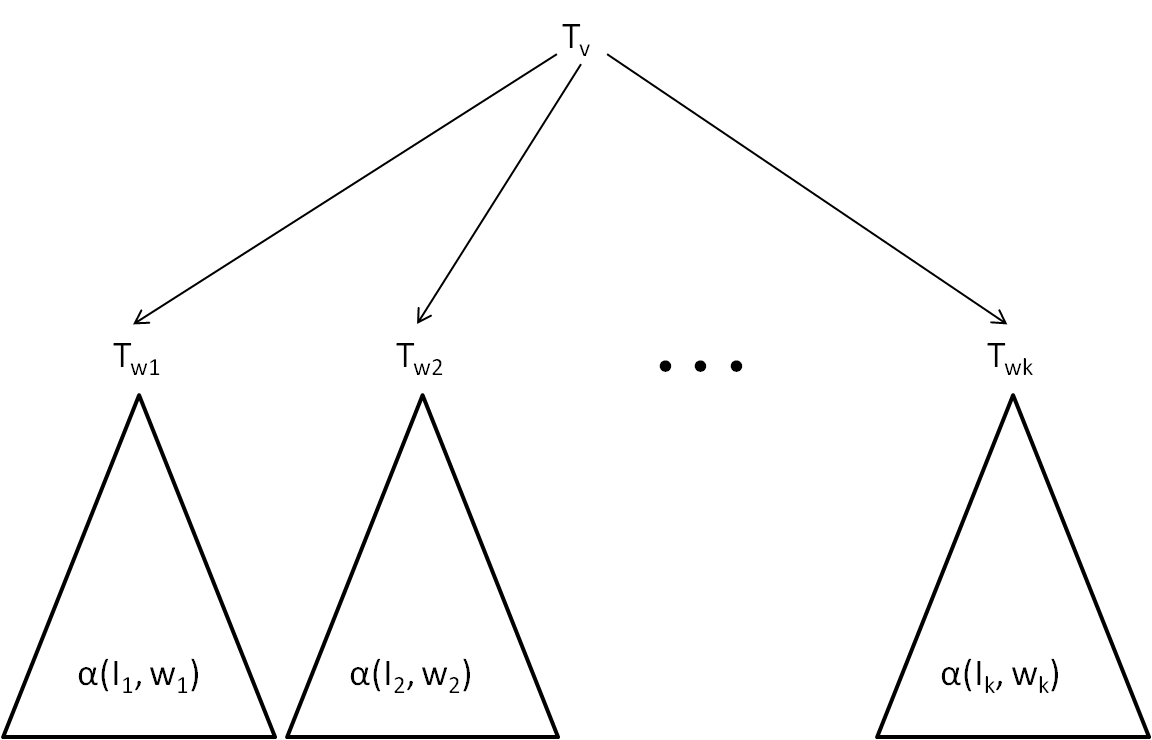}
\caption{The inner probabilities recurrence.}
\end{figure}

The \textbf{outer} probability $\beta({\mathbf I},v)$ is the probability of the entire tree excluded a subtree, rooted at particle of type $T_v$  and producing outcome ${\mathbf I} = \{i_1, i_2, \ldots, i_d\}$ (fig. 4). The recurrence here is:
\begin{flalign*}
& \beta({\mathbf I},v)  = \sum_w \sum_{\mathbf {v}} p(T_w \rightarrow \{T_v, T_{v_{(2)}}, \ldots, T_{v_{(m)}}\}) \\
& \times \sum_{{\mathbf J} \subset {\mathbf X} - {\mathbf I}} \beta({\mathbf I}+{\mathbf J}, w) \sum_{{\mathbf J_2}+\ldots+{\mathbf J_m}={\mathbf J}} \alpha({\mathbf J_2},v_{(2)})\ldots\alpha({\mathbf J_m},v_{(m)})
\end{flalign*}
where $\{T_v, \mathbf {v}\} = \{T_v, T_{v_{(2)}}, \ldots, T_{v_{(m)}}\}$ are all possible sets of particles that $T_w$ can produce and ${\mathbf X} = \{x_1, x_2, \ldots, x_d\}$ is the observed set of objects.

\begin{figure}[h]
\centering
\includegraphics[width=85mm]{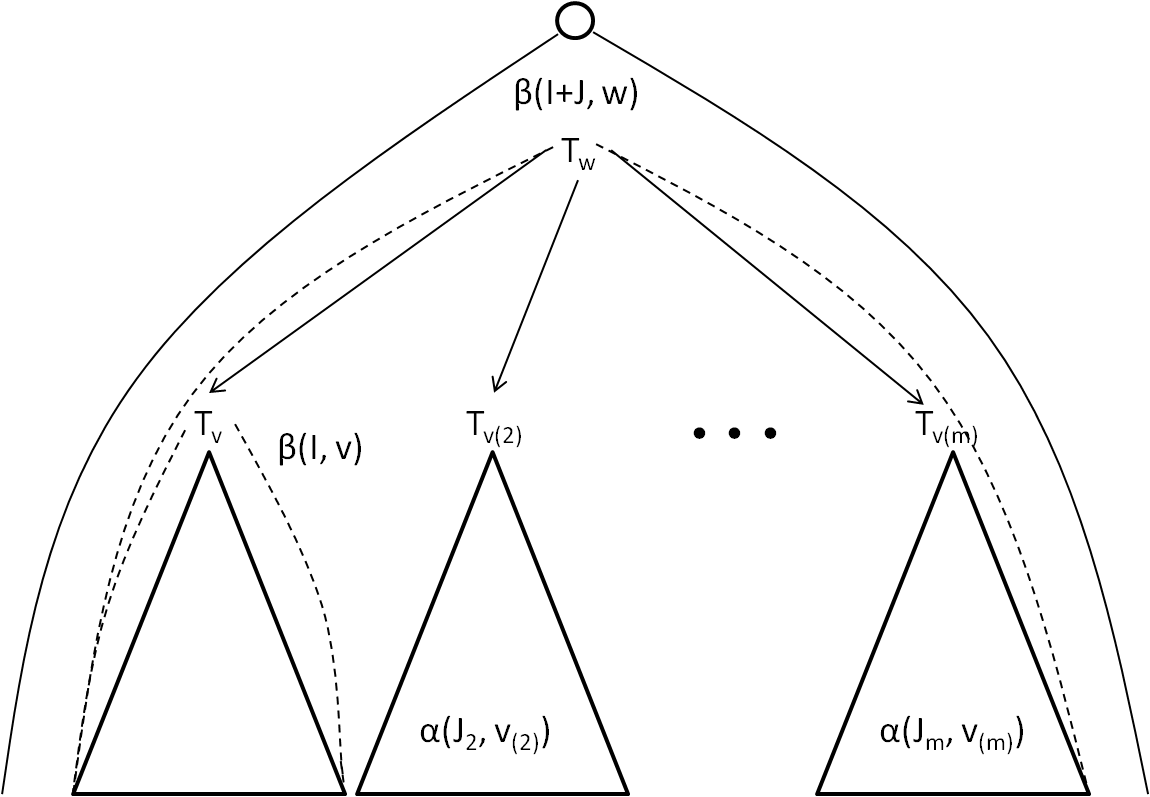}
\caption{The outer probabilities recurrence.}
\end{figure}

For every $T_v \in \pi$ the expected number of times $c(T_v)$ that $T_v$ is used in the tree $\pi$ could be presented as follows:
\begin{flalign*}
& E_{\mathbf{p}}c(T_v) = \sum_{\pi}P(\pi|x, \mathbf{p})c(T_v;\pi,x) = \sum_{\pi}\frac{P(\pi,x|\mathbf{p})}{P(x|\mathbf{p})}c(T_v;\pi,x)\\ 
& = \frac{1}{P(x|\mathbf{p})}\sum_{\pi}P(\pi,x|\mathbf{p})c(T_v;\pi,x) =  \frac{1}{P(x|\mathbf{p})}\sum_{\pi: T_v \in \pi}P(\pi,x|\mathbf{p})\\ 
& = \frac{1}{P(x|\mathbf{p})}\sum_{\mathbf I}\alpha({\mathbf I},v)\beta({\mathbf I},v),
\end{flalign*}
where $\alpha({\mathbf I},v)$ and $\beta({\mathbf I},v)$ are the inner and outer probabilities for all ${\mathbf I} \subset {\mathbf X}$.

Similarly, the expected number of times a production $T_v \rightarrow \{T_{w_1}, \ldots, T_{w_k}\}$ is used could be calculated:
\begin{flalign*}
& E_{\mathbf{p}^{(i)}}c(T_v \rightarrow \{T_{w_1}, \ldots, T_{w_k}\})\\
& = \sum\limits_{\mathbf I} \sum\limits_{{\mathbf I_1}+\ldots+{\mathbf I_k}={\mathbf I}} \beta({\mathbf I},v)\alpha({\mathbf I_1},w_1)\ldots\alpha({\mathbf I_k},w_k)p^{(i)}(T_v \rightarrow \{T_{w_1}, \ldots, T_{w_k}\}).
\end{flalign*}

Dividing the expectations above, we obtain the \textbf{EM re-estimation} of the parameters:
\begin{flalign*}
& p^{(i+1)}(T_v \rightarrow \{T_{w_1}, \ldots, T_{w_k}\}) \\
& = \frac{\sum\limits_{\mathbf I} \sum\limits_{{\mathbf I_1}+\ldots+{\mathbf I_k}={\mathbf I}} \beta({\mathbf I},v)\alpha({\mathbf I_1},w_1)\ldots\alpha({\mathbf I_k},w_k)p^{(i)}(T_v \rightarrow \{T_{w_1}, \ldots, T_{w_k}\})}{\sum\limits_{\mathbf I}\alpha({\mathbf I},v)\beta({\mathbf I},v)}
\end{flalign*}

For several observed sets of objects the expected numbers in the nominator and denominator are summed for all sets.

Such generally stated the algorithm still has high complexity. In practical applications though, often there are small number of types and not all of the productions in the offspring distributions are allowed. Thus, for a number of cases, a specific dynamic programming schemes based on the above recurrence could be proposed, which will be less complex.

\section{A Comprehensive Example}

We consider a MTBP with four types of particles -- two nonterminal $T_1$, $T_2$ and two terminal $T_1^T$, $T_2^T$, and the productions that are allowed (with nonzero probability) are:
\begin{equation}\label{ndaskalova:simmod}
T_1 {\stackrel{p^1_{11}}{\longrightarrow}} \{T_1, T_1\}, \quad T_1 {\stackrel{p^1_{12}}{\longrightarrow}} \{T_1, T_2\}, \quad T_1 {\stackrel{p^1_{T}}{\longrightarrow}} \{T_1^T\},\quad
T_2 {\stackrel{p^2_{22}}{\longrightarrow}} \{T_2, T_2\}, \quad T_2 {\stackrel{p^2_{T}}{\longrightarrow}} \{T_2^T\}.
\end{equation}
To cover the possibility of observing particles of types $T_1$ and $T_2$ the additional productions $T_1 {\stackrel{p^1}{\longrightarrow}} \{T_1\}$ and $T_2 {\stackrel{p^2}{\longrightarrow}} \{T_2\}$ are introduced.

Now, let $\mathbf{X} = \{1,0,1,1\}$ and the process has started with one particle of type $T_1$.
We don't have further information about the distribution, so an uniform one is assumed for every type: $p^1_{11} = p^1_{12} = p^1_{T} = p^1_{1} = 1/4$ and $p^2_{22} = p^2_{T} = p^2_{2} = 1/3$. 

Let $\alpha(\{i_1,i_2,i^T_1,i^T_2\}, v)$ be the probability of a tree rooted at a particle of type $T_v$ to produce the number of particles of types $T_1,T_2,T^T_1,T^T_2$ respectively. The initialization of $\alpha$ should be as follows:
\begin{equation*}
\alpha_{(1,0,0,0)}^1 = p^1_{1} = 1/4, \quad \alpha_{(0,0,1,0)}^1 = p^1_{T} = 1/4, \quad\alpha_{(0,1,0,0)}^2 = p^2_{2} = 1/3, \quad 
 \alpha_{(0,0,0,1)}^2 = p^2_{T}= 1/3.
\end{equation*}

For the level of two particles we are interested in all possible subsets of $\{1,0,1,1\}$ containing two $1$-s. The values of $\alpha$ are calculated below:
\begin{flalign*}
& \alpha_{(1,0,1,0)}^1 = p_{11}^1\alpha_{(1,0,0,0)}^1\alpha_{(0,0,1,0)}^1 + p_{12}^1\alpha_{(1,0,0,0)}^1\alpha_{(0,0,1,0)}^2 + p_{12}^1\alpha_{(0,0,1,0)}^1\alpha_{(1,0,0,0)}^2 \\
& = 1/4.1/4.1/4 + 1/4.1/4.0 + 1/4.1/4.0 = 1/64,
\end{flalign*}
Similarly, 
\begin{equation*}
\alpha_{(1,0,0,1)}^1 = 1/48 \quad \mbox{and} \quad \alpha_{(0,0,1,1)}^1 = 1/48.
\end{equation*}
Also
\begin{equation*}
\alpha_{(1,0,1,0)}^2 = 0,\quad \alpha_{(1,0,0,1)}^2 = 0, \quad \mbox{and} \quad \alpha_{(0,0,1,1)}^2 = 0.
\end{equation*}
Finally
\begin{flalign*}
& \alpha_{(1,0,1,1)}^1 = p_{11}^1[\alpha_{(1,0,1,0)}^1\alpha_{(0,0,0,1)}^1 + \alpha_{(1,0,0,1)}^1\alpha_{(0,0,1,0)}^1 + \alpha_{(1,0,0,0)}^1\alpha_{(0,0,1,1)}^1] + p_{12}^1[\alpha_{(1,0,1,0)}^1\alpha_{(0,0,0,1)}^2\\
& + \alpha_{(1,0,0,1)}^1\alpha_{(0,0,1,0)}^2 + \alpha_{(1,0,0,0)}^1\alpha_{(0,0,1,1)}^2 + \alpha_{(1,0,1,0)}^2\alpha_{(0,0,0,1)}^1 + \alpha_{(1,0,0,1)}^2\alpha_{(0,0,1,0)}^1 + \alpha_{(1,0,0,0)}^2\alpha_{(0,0,1,1)}^1] \\
& = 1/4.(1/64.0 + 1/48.1/4 + 1/4.1/48)+ 1/4.(1/64.1/3 + 1/48.0 + 1/4.0 + 0.0 + 0.1/4 + 0.1/48)\\
& = 1/256.
\end{flalign*}

Next follow the calculations of the outer probabilities $\beta(\{i_1,i_2,i^T_1,i^T_2\}, v)$. The initial values are $\beta(\{1,0,1,1\}, 1) = 1$ and $\beta(\{1,0,1,1\}, 2) = 0$. Then:
\begin{flalign*}
& \beta_{(1,0,1,0)}^1 = \beta_{(1,0,1,1)}^1[p_{11}^1\alpha_{(0,0,0,1)}^1 + p_{12}^1\alpha_{(0,0,0,1)}^2] = 1.(1/4.0 + 1/4.1/3) = 1/12,\\
& \beta_{(1,0,0,1)}^1 = 1/16, \quad \beta_{(0,0,1,1)}^1 = 1/16,\\
& \beta_{(1,0,1,0)}^2 = \beta_{(1,0,1,1)}^1p_{12}^1\alpha_{(0,0,0,1)}^1 + \beta_{(1,0,1,1)}^2p_{22}^2\alpha_{(0,0,0,1)}^2 {\small = 1.1/4.0 + 0.1/3.1/3 = 0}, \\
& \beta_{(1,0,0,1)}^2 = 1/16, \quad \beta_{(0,0,1,1)}^2 = 1/16,\\
\end{flalign*}
\begin{flalign*}
& \beta_{(1,0,0,0)}^1 = \beta_{(1,0,1,0)}^1[p_{11}^1\alpha_{(0,0,1,0)}^1 + p_{12}^1\alpha_{(0,0,1,0)}^2] \\
& + \beta_{(1,0,0,1)}^1[p_{11}^1\alpha_{(0,0,0,1)}^1 + p_{12}^1\alpha_{(0,0,0,1)}^2] + \beta_{(1,0,1,1)}^1[p_{11}^1\alpha_{(0,0,1,1)}^1 + p_{12}^1\alpha_{(0,0,1,1)}^2] \\ 
& = 1/12.(1/4.1/4 + 1/4.0) + 1/16.(1/4.0 + 1/4.1/3) + 1.(1/4.1/48 + 1/4.0) = 1/64, \\
& \beta_{(0,0,1,0)}^1 = 1/64 \quad \quad \mbox{and} \quad \quad \beta_{(0,0,0,1)}^1 = 3/256,
\end{flalign*}
\begin{flalign*}
& \beta_{(1,0,0,0)}^2 = \beta_{(1,0,1,0)}^1p_{12}^1\alpha_{(0,0,1,0)}^1 + \beta_{(1,0,1,0)}^2p_{22}^2\alpha_{(0,0,1,0)}^2 + \beta_{(1,0,0,1)}^1[p_{12}^1\alpha_{(0,0,0,1)}^1\\
& + \beta_{(1,0,0,1)}^2p_{22}^2\alpha_{(0,0,0,1)}^2] + \beta_{(1,0,1,1)}^1[p_{12}^1\alpha_{(0,0,1,1)}^1 + \beta_{(1,0,1,1)}^1p_{22}^2\alpha_{(0,0,1,1)}^2]\\ 
& = 1/12.1/4.1/4 + 0.1/3.0 + 1/16.1/4.0 + 1/16.1/3.1/3 + 1.1/4.1/48 + 1.1/3.0 = 5/288, \\
& \beta_{(0,0,1,0)}^2 = 5/288, \quad \quad \mbox{and} \quad \quad \beta_{(0,0,0,1)}^2 = 1/256. 
\end{flalign*}

Using that $P(x|\theta) = \alpha(\mathbf{X}, 1) = \alpha(\{1,0,1,1\}, 1) = 1/256$, we are able to compute the expected values we need:
\begin{flalign*}
& E_{\theta}c(T_1) = \frac{1}{P(x|\theta)}\sum_{\mathbf I}\alpha_{\mathbf I}^1\beta_{\mathbf I}^1 
= \frac{1}{1/256}(1/4.1/64 + 1/4.1/64 + 0 \\
& + 1/64.1/12 + 1/48.1/16 + 1/48.1/16 + 1/256) = \frac{4/256}{1/256} = 4,
\end{flalign*}
\begin{flalign*}
& E_{\theta}c(T_1 \rightarrow \{T_1, T_1\}) = \frac{1}{P(x|\theta)}\sum_{\mathbf I} \sum_{{\mathbf I_1}+{\mathbf I_2}={\mathbf I}} \beta_{\mathbf I}^1\alpha_{\mathbf I_1}^1\alpha_{\mathbf I_2}^1p(T_1 \rightarrow \{T_1, T_1\}) = \frac{1/256}{1/256} = 1 
\end{flalign*}
\begin{flalign*}
& E_{\theta}c(T_1 \rightarrow \{T_1, T_2\}) = \frac{1}{P(x|\theta)}\sum_{\mathbf I} \sum_{{\mathbf I_1}+{\mathbf I_2}={\mathbf I}} \beta_{\mathbf I}^1\alpha_{\mathbf I_1}^1\alpha_{\mathbf I_2}^2p(T_1 \rightarrow \{T_1, T_2\}) = \frac{1/256}{1/256} = 1 
\end{flalign*}
\begin{flalign*}
& E_{\theta}c(T_1 \rightarrow \{T_1^T\}) = \frac{1}{P(x|\theta)}\beta_{T_1^T}^1p(T_1 \rightarrow \{T_1^T\})  = \frac{1}{1/256}1/4.1/64 = \frac{1/256}{1/256} = 1,
\end{flalign*}
\begin{flalign*}
& E_{\theta}c(T_1 \rightarrow \{T_1\}) = \frac{1}{P(x|\theta)}\beta_{T_1}^1p(T_1 \rightarrow \{T_1\})  = \frac{1}{1/256}1/4.1/64 = \frac{1/256}{1/256} = 1.
\end{flalign*}
Thus, the estimated distribution for $T_1$ is 
\begin{flalign*}
& \hat{p}_{11}^1 = \frac{E_{\theta}c(T_1 \rightarrow \{T_1, T_1\})}{E_{\theta}c(T_1)} = 1/4, \quad 
 \hat{p}_{12}^1 = \frac{E_{\theta}c(T_1 \rightarrow \{T_1, T_2\})}{E_{\theta}c(T_1)} = 1/4, \\
& \hat{p}_{1}^1 = \frac{E_{\theta}c(T_1 \rightarrow \{T_1\})}{E_{\theta}c(T_1)} = 1/4, \quad
 \hat{p}_{T}^1 = \frac{E_{\theta}c(T_1 \rightarrow \{T_1^T\})}{E_{\theta}c(T_1)} = 1/4,
\end{flalign*}
which is the same as the initially chosen one, so this would be the final estimation.

For the offspring distribution of $T_2$ similar computations lead to the result:
{\small $$E_{\theta}c(T_2) = 1, \quad E_{\theta}c(T_2 \rightarrow \{T_2, T_2\}) = 0,  \quad E_{\theta}c(T_2 \rightarrow \{T_2\}) = 0,  \quad E_{\theta}c(T_2 \rightarrow \{T_2^T\}) = 1,$$}
so the estimation is:
\begin{align*}
\hat{p}_{22}^2 = 0, \quad \hat{p}_{2}^2 = 0, \quad \hat{p}_{T}^2 = 1,
\end{align*}
which converges on the next iteration also.

\section{Simulation study}

Simulation experiment has been performed to study behaviour of the estimates obtained via the algorithm. Observations have been simulated according to the model (\ref{ndaskalova:simmod}) in the previous section with offspring probabilities $p^1_{11} = p^1_{12} = p^1_{T} = 1/3$ and $p^2_{22} = p^2_{T} = 1/2$. Estimation has been performed using different sample sizes both for the number of observations, and the tree sizes as well. All the computations were made in R \cite{R}. 

It is important to investigate how the size of the tree, which corresponds to the size of the "hidden" part of the observation, affects the estimates. In Table \ref{ndaskalova:tabl1} are shown the result for small tree sizes and sample size 20. The most accurate estimates are obtained through averaging these results. It can be seen that there is great variation in the estimate of the individual distribution of type 2, thought the mean is close to the real values. For larger sample sizes the variance of the estimates is reduced, but there is some bias in the estimate for type 2 (Table \ref{ndaskalova:tabl2}). Larger sample trees also lead to biased estimates for the individual distribution for type 2 (Table \ref{ndaskalova:tabl3}). Using larger trees is also computationally more expensive. 

\begin{table}[h]
\begin{tabular}{|l|c|c|c|c|c|}
\hline size 20 & $p^1_{T}$ & $p^1_{11}$ & $p^1_{12}$ & $p^2_{T}$ & $p^2_{22}$\\
\hline s.1 & 0.35 & 0.38 & 0.27 & 0.25 & 0.75\\
\hline s.2 & 0.26 & 0.45 & 0.30 & 0.56 & 0.44\\
\hline s.3 & 0.33 & 0.35 & 0.32 & 0.47 & 0.53\\
\hline s.4 & 0.30 & 0.33 & 0.37 & 0.47 & 0.53\\
\hline s.5 & 0.33 & 0.38 & 0.28 & 0.00 & 1.00\\
\hline s.6 & 0.39 & 0.42 & 0.19 & 0.00 & 1.00\\
\hline s.7 & 0.38 & 0.35 & 0.27 & 0.94 & 0.06\\
\hline s.8 & 0.42 & 0.25 & 0.33 & 0.72 & 0.28\\
\hline 
\hline mean & 0.34 & 0.36 & 0.29 & 0.43 & 0.57\\
\hline st.dev. & 0.05 & 0.06 & 0.05 & 0.33 & 0.33\\
\hline
\end{tabular} 
\begin{tabular}{|l|c|c|c|c|c|} 
\hline size 20 & $p^1_{T}$ & $p^1_{11}$ & $p^1_{12}$ & $p^2_{T}$ & $p^2_{22}$\\
\hline s.9 & 0.34 & 0.39 & 0.27 & 0.00 & 1.00\\
\hline s.10 & 0.36 & 0.29 & 0.36 & 0.77 & 0.23\\
\hline s.11 & 0.31 & 0.29 & 0.40 & 0.46 & 0.54\\
\hline s.12 & 0.33 & 0.41 & 0.26 & 0.59 & 0.41\\
\hline s.13 & 0.48 & 0.17 & 0.36 & 0.92 & 0.08\\
\hline s.14 & 0.34 & 0.25 & 0.41 & 0.64 & 0.36\\
\hline s.15 & 0.20 & 0.33 & 0.47 & 0.93 & 0.07\\
\hline s.16 & 0.18 & 0.25 & 0.57 & 0.66 & 0.34\\
\hline 
\hline mean & 0.32 & 0.30 & 0.39 & 0.62 & 0.38\\
\hline st.dev. & 0.09 & 0.08 & 0.10 & 0.30 & 0.30\\
\hline 
\end{tabular} 

\centering
\begin{tabular}{|l|c|c|c|c|c|} 
\hline all samples & $p^1_{T}$ & $p^1_{11}$ & $p^1_{12}$ & $p^2_{T}$ & $p^2_{22}$\\
\hline avg. & 0.33 & 0.33 & 0.34 & 0.52 & 0.48\\
\hline st.dev. & 0.07 & 0.08 & 0.09 & 0.32 & 0.32\\
\hline 
\end{tabular} 
\caption{Estimation obtained using small tree samples of size 20.}
\label{ndaskalova:tabl1}
\end{table}

\begin{table}[h]
\begin{tabular}{|l|c|c|c|c|c|}
\hline size 50 & $p^1_{T}$ & $p^1_{11}$ & $p^1_{12}$ & $p^2_{T}$ & $p^2_{22}$\\
\hline s.1 & 0.29 & 0.40 & 0.31 & 0.39 & 0.61\\
\hline s.2 & 0.34 & 0.35 & 0.31 & 0.45 & 0.55\\
\hline s.3 & 0.36 & 0.36 & 0.27 & 0.82 & 0.18\\
\hline s.4 & 0.39 & 0.32 & 0.30 & 0.52 & 0.48\\
\hline s.5 & 0.31 & 0.35 & 0.34 & 0.55 & 0.45\\
\hline s.6 & 0.27 & 0.43 & 0.30 & 0.81 & 0.19\\
\hline 
\hline mean & 0.33 & 0.37 & 0.30 & 0.59 & 0.41\\
\hline st.dev. & 0.04 & 0.04 & 0.02 & 0.18 & 0.18\\
\hline 
\end{tabular} 
\begin{tabular}{|l|c|c|c|c|c|}
\hline size 100 & $p^1_{T}$ & $p^1_{11}$ & $p^1_{12}$ & $p^2_{T}$ & $p^2_{22}$\\
\hline s.1 & 0.31 & 0.38 & 0.31 & 0.41 & 0.59\\
\hline s.2 & 0.38 & 0.34 & 0.28 & 0.68 & 0.32\\
\hline s.3 & 0.29 & 0.39 & 0.31 & 0.67 & 0.33\\
\hline  
\hline mean & 0.33 & 0.37 & 0.30 & 0.59 & 0.41\\
\hline st.dev. & 0.04 & 0.03 & 0.02 & 0.16 & 0.16\\
\hline 
\end{tabular} 
\caption{Estimation obtained using small tree samples of size 50 and 100.}
\label{ndaskalova:tabl2}
\end{table}

\begin{table}[h]
\begin{tabular}{|l|c|c|c|c|c|}
\hline size 50 & $p^1_{T}$ & $p^1_{11}$ & $p^1_{12}$ & $p^2_{T}$ & $p^2_{22}$\\
\hline s.1 & 0.33 & 0.32 & 0.35 & 0.64 & 0.36\\
\hline s.2 & 0.33 & 0.30 & 0.36 & 0.72 & 0.28\\
\hline s.3 & 0.33 & 0.28 & 0.39 & 0.80 & 0.20\\
\hline s.4 & 0.34 & 0.33 & 0.33 & 0.68 & 0.32\\
\hline mean & 0.33 & 0.31 & 0.36 & 0.71 & 0.29\\
\hline st.dev. & 0.00 & 0.02 & 0.02 & 0.07 & 0.07\\
\hline 
\end{tabular} 
\begin{tabular}{|l|c|c|c|c|c|}
\hline size 100 & $p^1_{T}$ & $p^1_{11}$ & $p^1_{12}$ & $p^2_{T}$ & $p^2_{22}$\\
\hline s.1 & 0.34 & 0.31 & 0.36 & 0.71 & 0.29\\
\hline s.2 & 0.33 & 0.30 & 0.36 & 0.74 & 0.26\\
\hline mean & 0.34 & 0.31 & 0.36 & 0.73 & 0.27\\
\hline st.dev. & 0.00 & 0.00 & 0.00 & 0.02 & 0.02\\
\hline 
\end{tabular} 
\caption{Estimation obtained using larger tree samples of size 50 and 100.}
\label{ndaskalova:tabl3}
\end{table}

The bias in the estimate for type 2 is due to the greater uncertainty in the process for type 2: these particles could be generated by a particle of their own type, as well as, by a particle of type 1. For example, production of one particle of type 1 and two of type 2 could happen in two ways: once $T_1 \rightarrow \{T_1, T_2\}$ and then $T_2 \rightarrow \{T_2, T_2\}$, or twice $T_1 \rightarrow \{T_1, T_2\}$. So, in general, productions $T_1 \rightarrow \{T_1, T_2\}$ take part more often in the estimation than productions $T_2 \rightarrow \{T_2, T_2\}$. As the branching is hidden and all possible generations have to be taken in account, this results in underestimation of $p^2_{22}$ and some overestimation of $p^1_{12}$ when that hidden part gets larger. 

\section{Conclusion}

A general EM algorithm has been proposed to find ML estimation of the offspring probabilities of MTBP with terminal types when only an observation of the generation at given moment is available. The example presented shows that the algorithm is straightforward and convenient to apply for a particular model. Simulation study shows that better estimates are obtained using smaller samples. Such algorithms would be useful in biological models like cell proliferation, genetics, genomics, evolution, and wherever a model of MTBP with terminal types is suitable.



\bigskip 

\normalsize \noindent \sl
  \parbox[c]{220pt}{Nina Daskalova\\
 Sofia University "St.Kliment Ohridski",\\
  Faculty of Mathematics and Informatics,\\
   Sofia, Bulgaria,\\
    e-mail: {\tt ninad@fmi.uni-sofia.bg} }\\

\end{document}